\documentclass[twoside]{article} 
\usepackage[accepted]{aistats_stripped}

\usepackage{microtype}
\usepackage{booktabs}
\usepackage{soul}
\usepackage{amsmath,amsfonts,amssymb,amsbsy}
\usepackage{bbm}
\usepackage{url}
\usepackage{xcolor}
\usepackage{footnote}
\usepackage{graphicx} %
\usepackage{caption}
\usepackage{subcaption}
\usepackage{placeins} %
\usepackage[titletoc]{appendix} %
\usepackage{natbib} %
\usepackage{wrapfig}
\usepackage{listings}
\usepackage{verbatim} %
\usepackage{times}
\usepackage[colorlinks,linkcolor=black,citecolor=blue,urlcolor=blue,filecolor=blue,backref=page]{hyperref}

\usepackage{ifthen}
\usepackage{tikz,pgfplots}
\usetikzlibrary{matrix}
\usetikzlibrary{calc}

\def\figpdfdir{fig/} %
\def\figtikzdir{tikz/} %

\newcommand{\minput}[2][]{
\ifthenelse{\equal{#1}{pdf}}
	{ \includegraphics{\figpdfdir #2} }
	{ \tikzset{external/remake next} \tikzsetnextfilename{#2} \input{\figtikzdir #2} }
}

\usetikzlibrary{external}
\tikzset{external/system call={lualatex
	\tikzexternalcheckshellescape -halt-on-error -interaction=batchmode
	-jobname "\image" "\texsource"}}
\newcommand{\vc}[1] { \mathbf{#1} } %
\newcommand{\vs}[1] { \boldsymbol{#1} }
\newcommand{\tp}{\mathsf{T}}
\newcommand{\diag}[1] {\text{diag}\left(#1\right)}
\newcommand{\ti}[1] { \tilde{#1} } 
 
\newcommand{\tx}[1] { \text{#1} } 
\newcommand{\given} { \,|\, }

\renewcommand{\Re} {\mathbb{R}}
\newcommand{\indicator}[1] { \mathbbm{1} \left(#1\right) }

\newcommand{\Normal}[1] { \mathrm{N} {\left(#1\right)}  }

\newcommand{\InvGamma}[1] { \mathrm{Inv\text{-}Gamma} {\left(#1\right)} }

\newcommand{\halfCauchy}[1] { \mathrm{C}^+ {\left(#1\right)} }

\newcommand{\captionspace}{\vspace{-1.5em}}

\begin{document}

\twocolumn[

\aistatstitle{Iterative Supervised Principal Components}

\aistatsauthor{ Juho Piironen \And Aki Vehtari  }
\aistatsaddress{ {\tt juho.piironen@aalto.fi} \And {\tt aki.vehtari@aalto.fi} } 
\vspace{-1.6em}
\aistatsaddress{Helsinki Institute for Information Technology, HIIT \\ Department of Computer Science, Aalto University}

]

\begin{abstract}

In high-dimensional prediction problems, where the number of features may greatly exceed the number of training instances, fully Bayesian approach with a sparsifying prior is known to produce good results but is computationally challenging.
To alleviate this computational burden, we propose to use a preprocessing step where we first apply a dimension reduction to the original data to reduce the number of features to something that is computationally conveniently handled by Bayesian methods.
To do this, we propose a new dimension reduction technique, called iterative supervised principal components (ISPC), which combines variable screening and dimension reduction and can be considered as an extension to the existing technique of supervised principal components (SPCs).
Our empirical evaluations confirm that, although not foolproof, the proposed approach provides very good results on several microarray benchmark datasets with very affordable computation time, and can also be very useful for visualizing high-dimensional data.
\end{abstract}

\section{INTRODUCTION}

Inference in high-dimensional problems, where the number of features may greatly exceed the number of training instances, remains a topic of active research.
The frequentist approaches typically formulate the problem as an optimization task with a penalty that forces the solutions to be sparse, the most popular example being the Lasso~\citep{tibshirani1996}, but various others have also been proposed~\citep[e.g,][]{fan2001, zou2005, zou2006, candes2007}.
In the Bayesian literature, the dominant approach is to use a use sparsifying prior, such as the spike-and-slab~\citep{mitchell1988,george1993} or the horseshoe~\citep{carvalho2010}. %
Inference is typically carried out by using Markov chain Monte Carlo (MCMC), but also expectation-maximization (EM) based mode finding strategies have gained popularity recently~\citep{rockova2014,chang2016,bhadra2017b}.
Empirical evidence indicates that the Bayesian approach is more accurate \citep{polson2011,piironen2017b,piironen2017c,bhadra2017b} but is computationally expensive for large number of features, especially if MCMC is used for inference.

This paper studies a practical strategy for alleviating the computational burden related to the Bayesian inference in these problems via dimension reduction.
We investigate the following two-step procedure.
First, we perform a dimension reduction which reduces the number of features to something that is computationally conveniently handled by fully Bayesian methods.
Second, we perform the Bayesian model fitting using the reduced set of features with a sparsifying prior that will discover which of these new features are the most relevant.

Although not routinely used in the Bayesian workflow, this approach is certainly not new and has actually been very successful in empirical evaluations.
Most notably, this was the key idea behind the overall winners of the NIPS 2003 feature selection challenge~\citep{neal2006}, who used feature screening based on univariate significance tests and dimension reduction with principal component analysis (PCA) to reduce the dimensionality of the problem.
There have also been many other explorations on these ideas (with both Bayesian and non-Bayesian emphasis), such as the supervised PCA (SPCA) \citep{bair2006,yu2006}.
Especially the various screening approaches have proved to be promising and have received attention during the recent years~\citep{fan2008,song2015,mukhopadhyay2016,ahmed2017,chen2017}.

We propose a new method, called iterative supervised PCA (ISPCA) that combines screening and dimension reduction in such a way that the produced set of features aims to be maximally relevant for predicting the target variable.
The method is most closely related to the SPCA and could be considered as an extended version of it.

The main contributions of the paper are summarized as follows.
	We present a (non-trivial) extension to the original SPCA method. %
	Unlike the original formulation, our method is model independent and does not need cross-validation for estimating the screening parameter.
	We also show how to handle multiclass classification problems, which was not discussed by the original SPCA paper.
	Based on the empirical evaluation, our method is overall competitive with the PCA and SPCA, but sometimes yields better results when used for predictive model construction.
	When used for visualizing high-dimensional data using only a few features, our method consistently yields at least as good and sometimes considerably better results than PCA or SPCA, which makes it a useful tool for exploratory analysis.

We would like to point out that the two-step procedure discussed in this paper is not fully Bayesian as it uses the data twice: first when constructing the new feature representation and second time when fitting the predictive model.
Nevertheless, we are sometimes willing to relax the full Bayesian view in the pursuit for a scalable method that allows us to handle high-dimensional problems in a computationally feasible manner.

\section{BACKGROUND}

This section briefly reviews some background essential for understanding our method.

\subsection{Principal Components}
\label{sec:pca}

Assume we are given dataset with feature matrix ${\vc X\in \Re^{n\times D}}$ and target values $\vc y\in \Re^n$.
Throughout this paper will assume each column of $\vc X$ is standardized to have a zero mean and unit variance if not otherwise stated.
In linear dimension reduction we find a transformed set of features $\vc Z$ that are typically (but not necessarily) orthogonal and obtained by a linear projection of the original feature matrix onto a set of vectors $\vc W \in \Re^{D\times K}$
\begin{align}
	\vc Z = \vc X \vc W.
\label{eq:decomp}
\end{align}
Principal components analysis (PCA), where $\vc W$ consists of $K \le \min(n-1,\,D)$ first right singular vectors of $\vc X$, is a well-known example of such method and a natural choice for dimension reduction.

However, because PCA is an unsupervised technique, there is no guarantee that the projections onto the first $K$ principal components would result in an informative set of features $\vc Z$ regarding the prediction of $\vc y$.
For instance, suppose we would like perform linear regression of $\vc y$ onto some set of features and $\vs \beta_* \in \Re^D$ denotes the optimal coefficients in the original feature space~$\vc X$.
Now, if the number of features $D$ greatly exceeds the number of training instances $n$ and there is enough variation in $\vc X$ unrelated to $\vc y$,  it is possible that $\vs \beta_*$ does not even belong to the column space of $\vc W$, which means it is impossible to recover the optimal solution using the transformed set of features $\vc Z$.
Even if the optimal solution would be recoverable, the solution is not necessarily sparse in the new feature space (even if it was in the original space) which can make the learning more difficult and may require a large number of transformed features~$K$.

\subsection{Supervised Principal Components}
\label{sec:spca}

Supervised PCA (SPCA) \citep{bair2006} is a technique to alleviate problems with the standard unsupervised PCA.
SPCs are computed as follows:
\begin{enumerate}
	\item Compute the univariate scores $s_j=S(\vc x_j,\vc y)$ between each feature $\vc x_j$ and the target variable $\vc y$.
	\item Retain only those features with univariate score above some threshold $\gamma$, and compute the first (or first few) principal components of the reduced feature matrix~$\vc X_\gamma$.
\end{enumerate}
The score function $S(\vc x_j,\vc y)$ is generally taken to be the absolute univariate regression coefficient between $\vc x_j$ and $\vc y$ which is up to a constant the same as the (absolute) correlation between the two variables.
For determining an appropriate threshold $\gamma$, \cite{bair2006} proposed to use cross-validation for the final prediction model that utilizes the extracted features.

SPCA can written in the form~\eqref{eq:decomp} by padding the principal components of $\vc X_\gamma$ with zeros corresponding to the features that were screened out.
The benefit of SPCA compared to the standard PCA is that the screening step anticipates other sources of variation in $\vc X$ unrelated to the target variable $\vc y$, and thus the extracted features will typically be more related to the relevant variation.

Some problems still persist, however.
One is that the screening step ignores the uncertainty about the relevance of the features with univariate score $s_j$ less than $\gamma$.
Although for many datasets this does not appear to be harmful from predictive point of view, we would like a more principled approach for treating the remaining features than simply ignoring them since it is possible for a feature to be relevant even if its univariate score would be exactly zero (see example in Sec.~\ref{sec:toyexample}).
Secondly, choosing the thresholding parameter via cross-validation makes the construction of SPCs dependent of the model used for prediction.
This can make the procedure computationally expensive (especially if Bayesian model is used) and it would be conceptually more satisfactory to find model independent procedure for the dimension reduction.
A third issue is that the original formulation of \cite{bair2006} does not provide a way of handling classification problems with more than two class.
The next section discusses our proposed method that is inspired by the idea of SPCA but aims to provide a solution for all these problems.

\section{ITERATIVE SUPERVISED PRINCIPAL COMPONENTS}
\label{sec:ispca}

This section discusses our proposed method of iterative supervised PCA (ISPCA).
We shall first outline the algorithm and then discuss its properties, further ideas and implementational details in more detail.

\subsection{Outline of the Algorithm}
\label{sec:ispca_algorithm}

The algorithm consists of iterating the following steps $K$ times:
\begin{enumerate}
	\item Compute the univariate scores $s_j = S(\vc x_j, \vc y)$ for each feature $\vc x_j$.
	\item Retain only features with univariate score $s_j > \gamma$, and compute the first principal component $\vc v_\gamma$ of these features $\vc X_\gamma$. Choose $\gamma$ so that the projection of $\vc X_\gamma$ onto this vector $\vc z_\gamma = \vc X_\gamma \vc v_\gamma$ maximises the score $S(\vc z_\gamma, \vc y)$. Denote the extracted feature by $\vc z$.
	\item Subtract the variation explained by $\vc z$ from each column in $\vc X$ (including those that were screened out at step~2) as $\vc x_j' = \vc x_j - b_j\, \vc z$\, where $b_j = (\vc z^\tp \vc z)^{-1} (\vc x_j^\tp \vc z)$. This yields a modified feature matrix $\vc X'$.
	\item Set $\vc X \leftarrow \vc X'$ and go to step 1.
\end{enumerate}

The intuition behind the algorithm is as follows.
At step~2 of each iteration, we seek direction that is maximally relevant for explaining variance of $\vc y$.
Step~3 ensures that the subsequent directions will capture variation that is not explained by the directions that we have computed so far.
This is useful, because there may be features in $\vc X$ that are screened out at step~2 but are still correlated with those that are retained after screening, and we do not want subsequent latent features $\vc z$ to be correlated (see discussion below).

\subsection{Properties of the Method}

Like PCA and SPCA, the algorithm in Section~\ref{sec:ispca_algorithm} results in a transformed set of features $\vc Z \in \Re^{n\times K}$ (computed at step~2) that are orthogonal and obtained by a linear projection of the original feature matrix onto a set of vectors $\vc W \in \Re^{D\times K}$ as in Equation~\eqref{eq:decomp}. %
To see that the features $\vc Z$ will be orthogonal, consider the following. %
After step 3 in the algorithm, all the columns of $\vc X'$ will be orthogonal to $\vc z$ because
\begin{align*}
	\vc x_j'^\tp \vc z 
	&= (\vc x_j - b_j \,\vc z)^\tp \vc z \\
	&= \left(\vc x_j - (\vc z^\tp \vc z)^{-1} (\vc x_j^\tp \vc z)\, \vc z \right)^\tp \vc z \\
	&= \vc x_j^\tp \vc z - (\vc z^\tp \vc z)^{-1} (\vc x_j^\tp \vc z)\, \vc z^\tp \vc z \\
	&= 0,
\end{align*}
and therefore any linear combination of these (that is, the latent feature to be extracted at the next iteration) will also be orthogonal to $\vc z$.
Using induction, it is straightforward to show that each latent feature is orthogonal also to all the other extracted features, not only to the previous one (the proof is omitted).
To prove that the extracted features can be written in the form of Equation~\eqref{eq:decomp}, we need to set up some notation.
Denote the feature matrix used at steps 1 and 2 at iteration~$k$ as $\vc X_k$, so that $\vc X_1 = \vc X$ is the original feature matrix.
Moreover, denote the principal components computed at step 2 as $\vc v_1,\dots, \vc v_K$. %
For notational convenience, we shall now assume that these vectors are padded with zeros corresponding to those features that were screened out at the corresponding iteration, so that each $\vc v_k \in \Re^D$.

Using this notation, the latent variables $\vc z_k$ computed at step~2 satisfy $\vc z_k = \vc X_k \vc v_k$ for all $k = 1,\dots,K$.
The construction of the next feature matrix $\vc X_{k+1}$ from the previous one $\vc X_k$ at step 3 can be written in a matrix form as
\begin{align}
	\vc X_{k+1} = \vc X_k - \vc Z_k \vc B_k,
\label{eq:X_next}
\end{align}
where all the columns of $\vc Z_k \in \Re^{n\times D}$ are equal to $\vc z_k$, and $\vc B_k$ is a diagonal matrix with elements $b_1,\dots,b_D$ from iteration~$k$. %
We can rewrite $\vc Z_k = \vc X_k \vc V_k$ where ${\vc V_k \in \Re^{D\times D}}$ with all columns equal to $\vc v_k$.
By plugging this into~\eqref{eq:X_next} we get
\begin{align*}
	\vc X_{k+1} &= \vc X_k - \vc X_k \vc V_k \vc B_k \\
				&= \vc X_k (\vc I - \vc V_k \vc B_k) \qquad | \quad \vc A_k := \vc I - \vc V_k \vc B_k \\
				&= \vc X_k \vc A_k, 
\end{align*}
from which we deduce
\begin{align*}
	\vc X_k = \vc X \vc A_1 \vc A_2 \dots \vc A_{k-1} =  \vc X \prod_{t=1}^{k-1} \vc A_t.
\end{align*}
This lets us write the latent features $\vc z_k$ as
\begin{align*}
\vc z_k = \vc X_k \vc v_k 
		= \vc X \left( \prod_{t=1}^{k-1} \vc A_t \right) \vc v_k,
\end{align*}
and thereby we arrive at decomposition~\eqref{eq:decomp} by defining the columns of the projection matrix $\vc W$ as
\begin{align}
	\vc w_k = \left( \prod_{t=1}^{k-1} \vc A_t \right) \vc v_k %
			= \left( \prod_{t=1}^{k-1} (\vc I - \vc V_t \vc B_t) \right) \vc v_k.
\label{eq:from_v_to_w}
\end{align}
In practice we never form matrices $\vc V_k$ or $\vc B_k$ to compute $\vc W$.
By exploiting the structure of these matrices, the columns of $\vc W$ can be computed much more efficiently (see details in the supplementary material).

It is worth noticing that although the new features $\vc Z$ will be orthogonal, the columns of the rotation matrix $\vc W$ typically will not.
This is not a handicap and can, in fact, be very beneficial as it allows detecting features that are not relevant alone but become relevant after some other features are included in the model (see Sec.~\ref{sec:toyexample} for a simple example).

\subsection{Combination of Supervised and Unsupervised Components }
\label{sec:combining_sup_unsup}

In principle, we could extract $\min(n-1,\,D)$ ISPCs from the data (or until none of the features have univariate score numerically distinguishable from zero).
In practice, however, this is not advisable and we call this the {\it naive} algorithm.
This is because the process of repeatedly finding the most relevant direction can overfit especially when the sample size $n$ is small because some features may have a relatively large absolute sample correlation with $\vc y$ although they are completely irrelevant, simply due to random fluctuation in the data.
Thus the algorithm may find ``relevant'' features that are in fact noise.
This will result in biases in the inference when the extracted features are later used for visualization or predictive model construction.
Thus in practice we typically extract only a few supervised components, and if needed, compute the standard unsupervised PCs with the rest of the data variation.
A practical automatic strategy for deciding the number of supervised components is discussed in Section~\ref{sec:permtest}.

After the supervised iteration, we can compute standard principal components as usual but with the exception that these are now computed from the modified data matrix $\vc X'$ that we are left with after the supervised iteration (after subtracting the variation explained by the $K$ supervised components at step 3 of each iteration).
If we denote the total number of components by $K_\tx{tot}$ and the unsupervised components by $\vc v_k,\,\, k=K+1,\dots,K_\tx{tot}$, the columns of the final projection matrix $\vc W \in \Re^{D\times K_\tx{tot}}$ corresponding to the unsupervised components are given by 
\begin{align}
	\vc w_k = \left( \prod_{t=1}^{K} (\vc I - \vc V_t \vc B_t) \right) \vc v_k, \quad k = K+1,\dots,K_\tx{tot}.
\end{align}
It is worth noticing that after this process, all the extracted features $\vc z_k$ (both supervised and unsupervised) will be orthogonal, which is often useful.
The inclusion of unsupervised components can be important for constructing a good predictive model.
This is simply due to the fact that not always all the relevant variation will be captured by the first supervised components.
This point will be demonstrated experimentally in Section~\ref{sec:experiments}.
We also point out that we can apply this same idea for the original SPCA, that is, compute unsupervised components from the features screened out, and make these unsupervised features orthogonal to the supervised ones.

\subsection{Deciding the Number of Supervised Components}
\label{sec:permtest}

As discussed in Section~\ref{sec:combining_sup_unsup}, the unrestricted supervised iteration may overfit, that is, find features that are appear relevant but are in fact noise.
Fortunately, there is a simple but effective way of discovering how many components we can extract without substantial overfitting.
We do this using a permutation test.
Before computing the next supervised principal component at step 2 of each iteration, we compute a $p$-value
\begin{align}
	p = 
	\frac{1}{R}\sum_{r=1}^R 
	\indicator{ \max_j S(\vc x_j, \vc y_r) \ge \max_j S(\vc x_j, \vc y) },
\label{eq:perm_pval}
\end{align}
where $\vc y_r$ denotes a random permutation of the original~$\vc y$ and $\indicator{E} = 1$ if event $E$ is true and zero otherwise.
Quantity~\eqref{eq:perm_pval} estimates how likely it is that the maximal univariate score would be as extreme as actually observed if none of the variables $\vc x_j$ were actually related to $\vc y$.
If $p < \alpha$ for a relatively small $\alpha$, we have strong evidence that there is still relevant variation left in the data and we can extract the next component being fairly confident that the finding was not a false discovery.
If $p \ge \alpha$, we stop the supervised iteration and proceed to extracting unsupervised features if needed.

In our experiments we used $\alpha= 0.01$ and $R=1000$ random permutations, which makes the number of false discoveries small. %
In principle we believe that it is better to be too conservative in setting $\alpha$ than to allow the algorithm to overfit.
After all, for predictive model construction we can always compute the standard unsupervised PCs with the rest of the data variation, use Bayesian model with a sparsifying prior and let the data decide which components are really relevant and which not.
The results indicate that this strategy is both computationally feasible and performs well in practice.

\subsection{More Algorithmic Details}

Finding the optimal screening threshold $\gamma$ at step 2 would require computing the first principal component for all feature subset sizes from 1 to $D$ which is computationally expensive.
In practice we use a more crude search and set up an evenly spaced  grid of values between $\gamma_\tx{min}$ and $\gamma_\tx{max}$, where $\gamma_\tx{max}$ is the smallest $\gamma$ so that all but one feature are screened out, and $\gamma_\tx{min}$ the largest $\gamma$ so that the number of features after screening is $W$.
We could set $W=D$ but since in practice the optimal $\gamma$ is rarely so that almost all features survive the screening, we typically use $W < D$ which makes the algorithm faster and concentrates the grid on more plausible values.
The computational complexity of computing a PC among at most $W$ features is the minimum of $O(W N^2)$ and $O(W^2 N)$, which shows that computational savings can be obtained by adjusting the feature window size $W$.
In our experiments we used grid of size $M = 10$ with feature window limit $W=500$ which seem to provide good balance between accuracy and speed.
In fact, especially when $n$ is fairly large, typically more time is spent in the permutation test (Sec.~\ref{sec:permtest}) which scales as $O(RND)$.

As a minor detail, we mention that before using the new features $\vc Z$ for predictive model construction, we typically normalize them to have unit variance so that none of the features is favored a priori.
For visualization purposes this is not necessary but does not hurt either.

\subsection{Multiclass Classification and Other Observation Models}

The supervised algorithm in Section~\ref{sec:ispca_algorithm} can naturally be extended to classification problems with $C > 2$ classes.
We do this by defining $C$ binary variables $\vc y_c=\indicator{\vc y=c}$, that is, ``class $c$ or some other class'', for $c=1,\dots,C$.
We then repeat steps 1 and 2 for all these $C$ auxiliary target variables which yields candidate directions $\vc v_c, \, c=1,\dots, C$ from which then choose the one which maximizes the score $S(\vc X \vc v_c, \vc y_c)$.
This typically results in direction $\vc v$ that tries to separate one of the classes from the rest (see the multiclass example in Sec.~\ref{sec:visualization}). 

We propose to use this same idea also for SPCA; in this case we define the univariate scores in the screening to be the maxima of the $C$ scores as $\ti S(\vc x_j,\vc y) = \max_c S(\vc x_j,\vc y_c)$.
Although simple, this approach turns out to be quite successful, and lets us extend also SPCA to multiclass problems (not discussed by \cite{bair2006}).

When computing the univariate scores $S(\vc x_j, \vc y)$ in other than regression or classification problems we could use pseudo-data $\vc t$ in place of $\vc y$, so that $\vc t$ is derived from the second order expansion to the likelihood from an univariate (generalized) regression of $\vc y$ onto $\vc x_j$.
This is discussed by \cite{bair2006} so we do not discuss it further here.

\begin{figure}[t]
	\centering
	\minput[pdf]{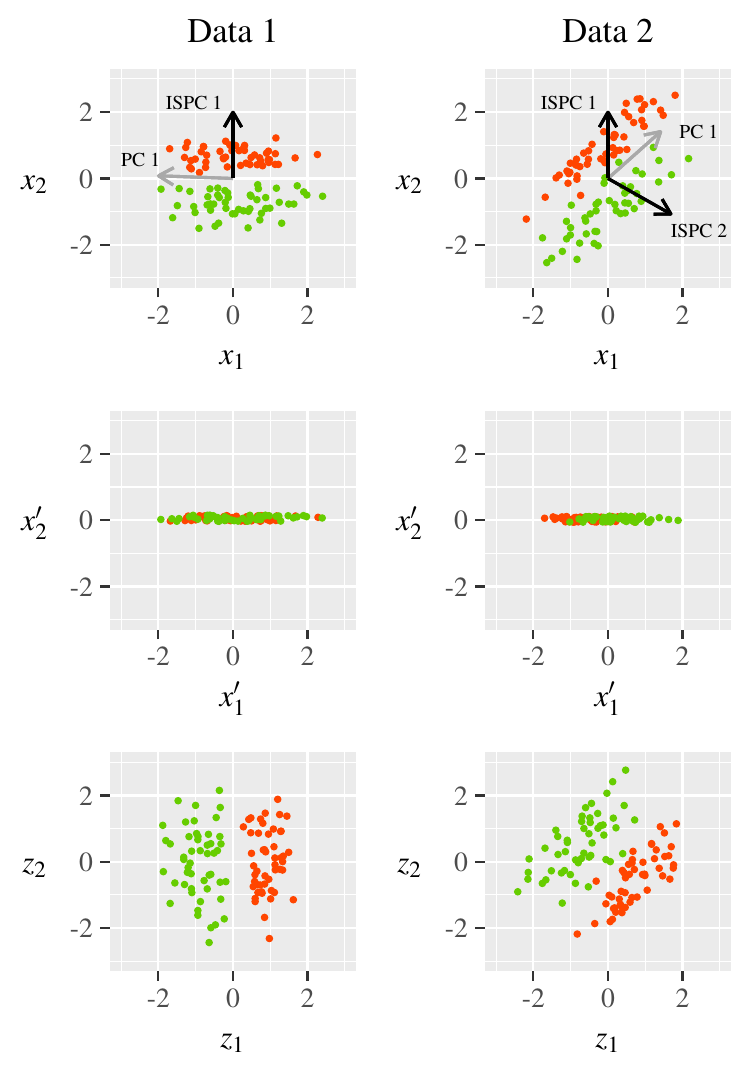}
	\captionspace
	\caption{Illustration of ISPCA for two toy binary classification datasets (the two columns). Top row shows the original dataset (colors denoting the different classes), supervised components found by ISPCA with the permutation test, and the first unsupervised PC. Middle row shows the new feature matrix obtained after subtracting the variation related to the first ISPC from $\vc X$ (in the plots, $x_2'$ is exactly zero but has been jittered by a small amount to aid visualization). Bottom row shows the the transformed features $\vc Z$.}
	\label{fig:toy}
\end{figure}
\begin{figure*}[t]
	\centering
	\minput[pdf]{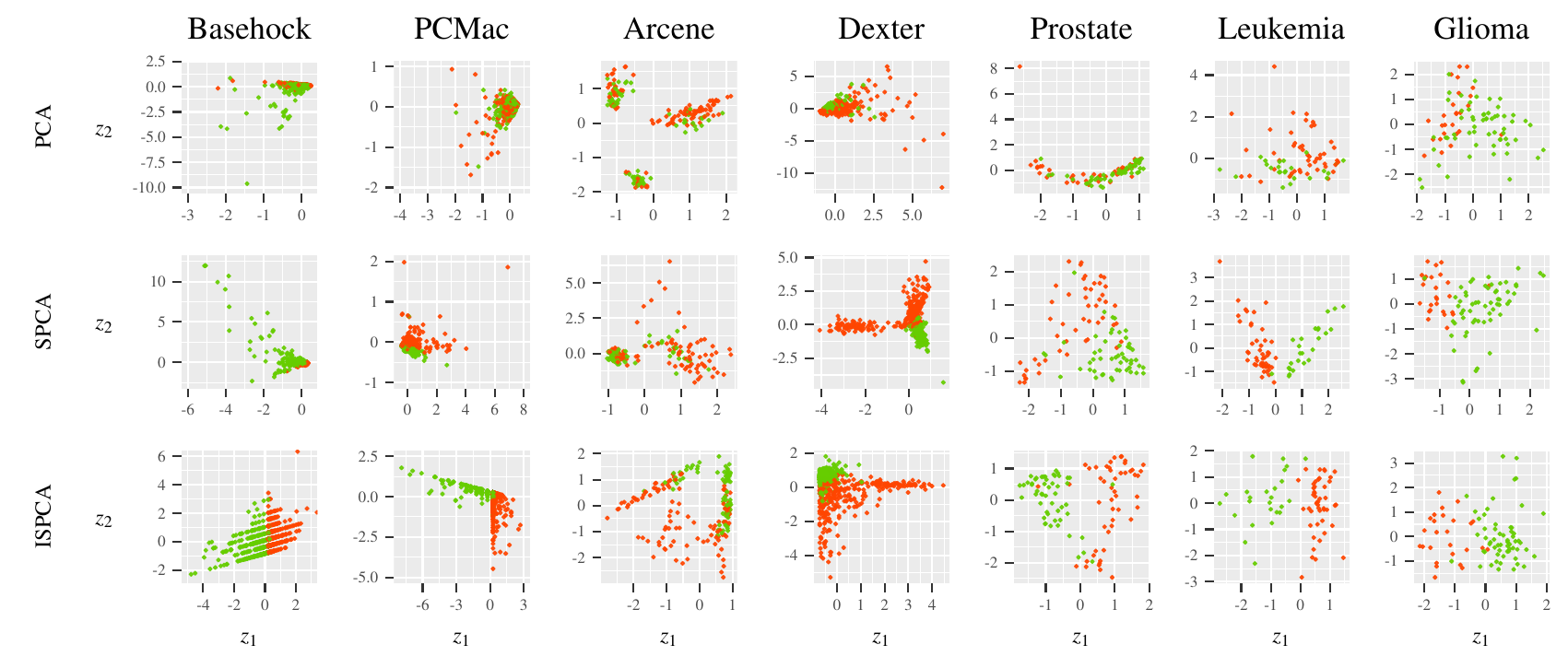}
	\captionspace
	\caption{Some of the binary classification datasets visualized using first two latent features obtained by PCA (top row), SPCA (middle row) and ISPCA (bottom row). Colors refer to the two classes. For Basehock and PCMac the visualization is done using only a subset of the data to reduce data overlap, but the features are extracted using the full datasets. For the last three datasets, the second feature of ISPCA is actually unsupervised, since only one supervised component was supported by the data.}
	\label{fig:datavis2c}
\end{figure*}

\subsection{Interpretation and Obtaining a Sparse Solution in the Original Space}

The columns of the projection matrix $\vc W$ are directly interpretable by investigating which entries are nonzero, as the corresponding features are likely to be correlated and predictive about $\vc y$ (at least if the column was computed in a supervised fashion).
If a linear model is used with the new set of features $\vc Z$, the corresponding regression coefficients $\vs {\ti \beta}$ can be transformed back to the original feature space simply as $\vs \beta = \vc W \vs {\ti \beta}$.

Although the columns of $\vc W$ will typically have a lot of zeros, they can still contain quite a few nonzeros as the nonzero entries correspond to correlated features that carry similar information.
To obtain an even more sparse solution in the original feature space, we can use the projective variable selection framework \citep{goutis1998,dupuis2003} which has shown to be successful for finding a sparse solution when there is redundancy in the features \citep{piironen2017a}.
This technique has also been studied from a non-Bayesian viewpoint with good results, and is known as ``preconditioning'' for variable selection \citep{paul2008}.
Due to the space constraints, we do not discuss this further but merely point out that this is possible.

\section{EXPERIMENTS}
\label{sec:experiments}

\subsection{Toy Examples}
\label{sec:toyexample}

We first illustrate the use of ISPCA with two simple toy problems that will shed light on the algorithm, see Figure~\ref{fig:toy}.
The first column shows data where only one of the variables ($x_2$) is relevant for separating the two classes.
Out of the two variables, $x_2$ has higher univariate score, and since this is higher than the univariate score for the principal component of the two features $(x_1,x_2)$, the first ISPC points to direction $\vc w=(0,1)$.
After subtracting the variation explained by this direction from the feature matrix $\vc X$, we end up with a modified feature matrix $\vc X'$ where only $x_1'$ has nonzero variance (middle row).
However, this feature has univariate score close to zero (thereby failing the permutation test), and hence the supervised iteration terminates.
If we compute unsupervised PCA using this rest of the data variation $\vc X'$, we end up with transformed features $(z_1,z_2)$ (bottom row), where only the first one is supervised and also the only relevant feature.

The second column shows a more interesting example.
Again, feature $x_1$ is irrelevant alone (has univariate score close to zero), but becomes relevant together with $x_2$ (that is, $x_1$ and $x_2$ together have better class separation than $x_2$ alone).
Again the first ISPC points towards $\vc w=(0,1)$, but now after subtracting the variance explained by this direction from $\vc X$, since $x_1$ and $x_2$ are correlated, we end up with a new feature matrix $\vc X'$ where the feature $x_1'$ has a significant correlation with the class label.
The first PC of $\vc X'$ points to direction $\vc v = (0,1)$, but transforming this back to the original feature space using Equation~\eqref{eq:from_v_to_w}, the second ISPC points roughly to direction $\vc w = (1.6,\, -1)$ in the original space (top plot).
This results in a new set of features $(z_1,z_2)$ out of which both are supervised and about equally relevant (bottom plot).

In both of these examples the first unsupervised PC does not explain variation relevant for separating the two classes.
SPCA would work well in the first case because then $x_1$ would be screened out and the first SPC would be equal to the first ISPC.
However, the second case shows an example where ISPCA has a distinctive advantage over the SPCA.
Also in this case SPCA would screen $x_1$ out and would find only the first relevant direction (that is, feature $x_2$), whereas the iterative procedure can discover that $x_1$ becomes relevant when $x_2$ is included.
Obviously, one could set the screening threshold $\gamma$ in SPCA so low that also $x_1$ would survive the screening, but in practice this means setting the threshold so low that basically all features are included, meaning that SPCA would in essence reduce to the standard PCA with the problems explained in Section~\ref{sec:pca}.

\subsection{Data Visualization}
\label{sec:visualization}

This section illustrates the use of ISPCA for visualization of high-dimensional real world data and shows how it compares to PCA and SPCA\footnote{Unlike in \cite{bair2006} who used a model dependent cross-validation scheme to chose the screening threshold for SPCA, we used here a simpler strategy and computed the $p$-values for each feature based on a permutation test for the univariate scores and retained only features with $p < 0.001$.}.
All the datasets involve a classification problem with the number of features ranging from about 1500 to 22000 and the number of training instances from about 50 to 2000.
Our main interest are the ``small~$n$, large $D$'' cases and most of the problems fall into this category, but we included also a few text classification problems.
See Table~\ref{tab:datasets} and the associated text in the supplementary material for more information about the datasets.
Figure~\ref{fig:datavis2c} shows a representative set of the binary classification datasets visualized using the first two latent features obtained using the three methods. 
The benefit of supervision for visualization purposes is very clear: in many cases the two classes are considerably overlapping when visualized using the first two unsupervised PCs, but become fairly well separated when using either SPCA or ISPCA.
By visual inspection, ISPCA seems to work clearly better than SPCA in at least one example (Basehock), slightly better in a few of the cases (PCMac, Arcene and Prostate) and slightly worse in one (Dexter).
\begin{figure}[t]
	\centering
	\minput[pdf]{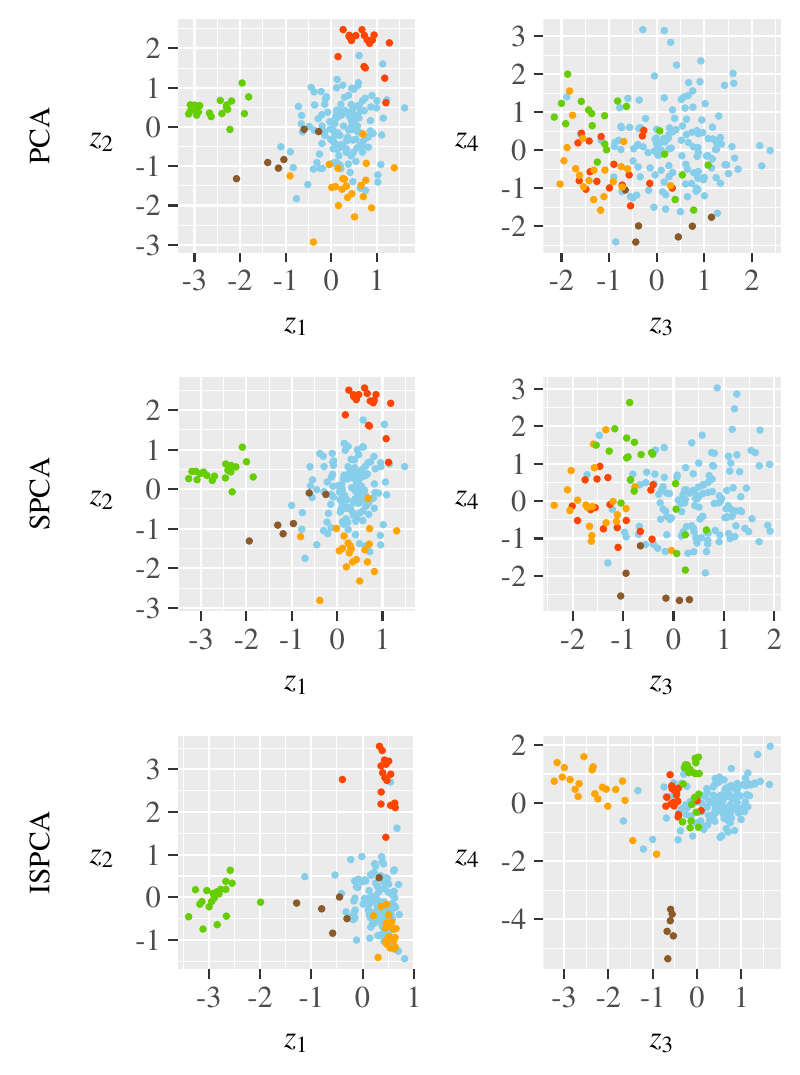}
	\captionspace
	\caption{Visualization of Lung-5c cancer data (${n=203, D=3312}$) using the first four latent features from PCA, SPCA and ISPCA. Different colours refer to the five different classes. Using only four features, {ISPCA} is able to separate the classes almost perfectly.}
	\label{fig:lung5c}
\end{figure}

Figure~\ref{fig:lung5c} shows a visualization of a dataset with five classes using the first four latent features of the three methods.
Here PCA and SPCA perform very similarly; the first two latent features are informative for separating the red and green classes from the rest, but the remaining two features are only weakly informative and in this plot the classes are considerably overlapping.
ISPCA on the other hand shows a substantial improvement; the method is able to find the third and fourth features so that also the orange and brown classes become well separated from the rest, and improves also the separation of the red and green class from the rest.
By investigating which of the entries in the vectors $\vc w_1,\dots,\vc w_4$ are nonzero we can get an idea about which of the features characterize the differences between the five classes (see Figure~\ref{fig:lung5c_ws} in the supplementary material).

\subsection{Predictive Model Construction}
\label{sec:prediction}

\begin{figure*}[t]
	\centering
	\minput[pdf]{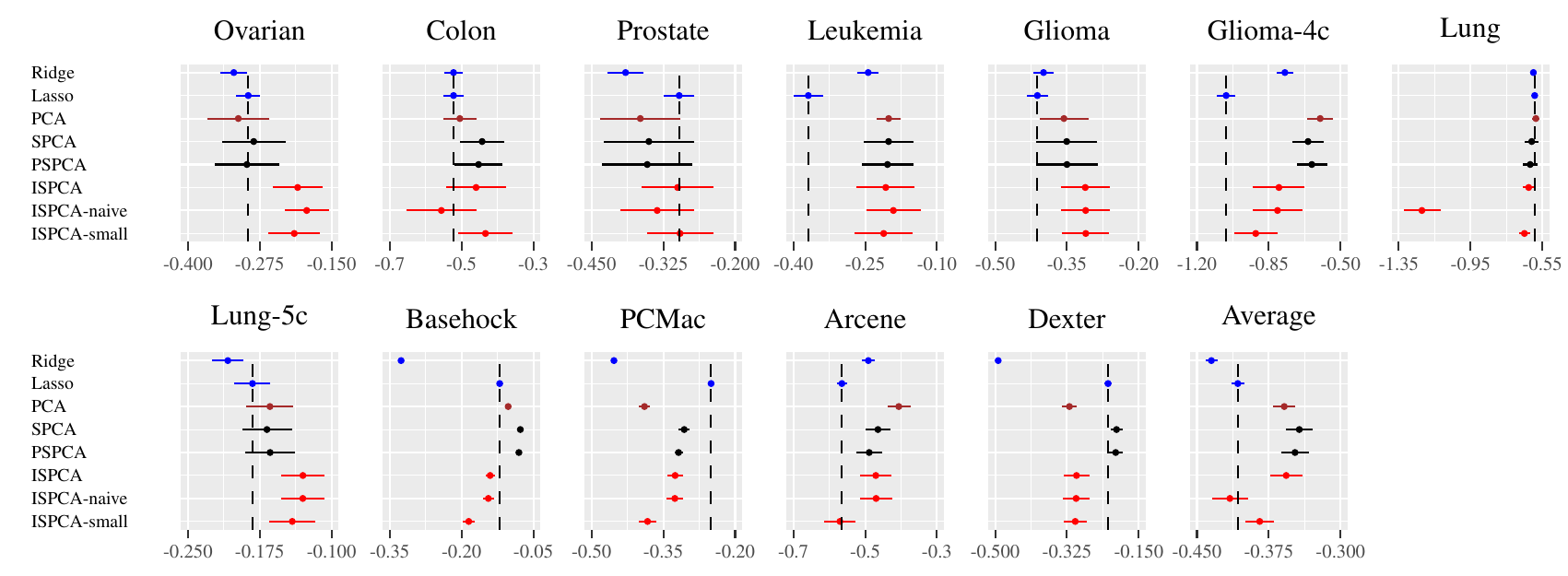}
	\captionspace
	\caption{Mean log predictive densities (MLPD) on test data for the different methods on different datasets (larger is better). Horizontal bars denote the 95\% intervals. The dashed vertical line denotes the performance estimate for the Lasso which was chosen as the baseline for the comparison. The last plot denotes the average over all the datasets.}
	\label{fig:mlpd}
\end{figure*}

Finally, we shall consider how the different dimension reduction techniques perform when the extracted latent features are used for predictive model construction.
In all cases we use the standard logistic regression model (in multiclass cases multinomial softmax regression) with the regularized horseshoe prior \citep{piironen2017c} for the regression coefficients.
This prior shrinks heavily towards zero the coefficients of irrelevant features and softly regularizes the coefficients of the relevant features.
More details of the prior, implementation of the models and computation can be found from the supplementary material.

We tested the following dimension reduction methods:
\begin{itemize}
	\item PCA: first $K_\tx{tot}$ unsupervised PCs ($K_\tx{tot}$ defined below).
	\item SPCA: supervised PCA, compute $K_\tx{tot}$ first PCs among those features  choose all features with  univariate score statistically significant ($p < 0.001$) 
	\item PSPCA: ``partially supervised PCA'', that is, compute first $K_\tx{tot}/2$ PCs among those features with  univariate score statistically significant, and compute first $K_\tx{tot}/2$ PCs using the rest of the features after subtracting the variation explained by the supervised components (see Sec.~\ref{sec:combining_sup_unsup}).
	\item ISPCA-naive: first $K_\tx{tot}$ ISPCs.
	\item ISPCA-small: first $K$ ISPCs, decide $K$ using the permutation test (Sec.~\ref{sec:permtest}).
	\item ISPCA: as ISPCA-small, but in addition compute $K_\tx{tot}-K$ unsupervised components.
\end{itemize}
For the binary classification datasets we used $K_\tx{tot}=50$ and multiclass problems $K_\tx{tot}=20$.
In addition we computed results also for ridge logistic regression and Lasso using the original features to get baseline results for comparisons.
The prediction accuracy was measured by splitting the data randomly into two parts, using one fifth as a test set, and then averaging the results over fifty such random splits.

Figure~\ref{fig:mlpd} shows the mean log predictive densities on test data for different methods on each dataset, the last plot denoting the average over all datasets (for classification accuracies, see Figure~\ref{fig:acc} in the supplementary material).
The results show that ISPCA yields better results than PCA or SPCA for several datasets, but also loses to one of these in many cases, the overall result being very close with SPCA having a slight edge (see the last plot).
Overall the dimension reduction techniques outperform Lasso by a clear margin, but the best method depends on data, which emphasizes that no single method is optimal for every problem.
The trend seems to be that ISPCA performs best on average for the microarray datasets (Ovarian -- Lung-5c) whereas SPCA works better for the text classification datasets where the features are word counts (Basehock, PCMac, Dexter).
This is an interesting pattern since it is somewhat at odds with the very good two dimensional feature representation of ISPCA which is better than for PCA and SPCA at least for Basehock and PCMac datasets (Figure~\ref{fig:datavis2c}).
These results call for further investigation to better understand the successes and failures of each method.

The comparison of ISPCA to ISPCA-naive and ISPCA-small show that not using the permutation test can lead to inferior results due to overfitting (Colon, Prostate, Lung-2c), and that inclusion of the unsupervised features is basically never harmful, but can clearly improve the results in some cases (PCMac, Arcene).
For SPCA the inclusion of unsupervised features does not appear to be crucial as the results for SPCA and PSPCA are practically the same for every dataset.

Computationally the three dimension reduction approaches are quite similar in the ``small $n$, large $D$'' realm, ISPCA being somewhat more expensive for multiclass problems and large $n$ (see Table~\ref{tab:computation_times} in the supplementary).
In these cases the number of supervised iterations is typically larger, which results in more permutation tests which are the most time consuming part of the method.
However, even in these cases the computational bottleneck is still the fitting of the predictive model.
Overall, although the considered methods cannot compete with Lasso in speed, the computation times are very affordable, considering that most problems allow the model to be fitted in a matter of seconds.

\section{CONCLUSIONS}

This paper has proposed a new supervised dimension reduction technique.
Our experiments indicate that the proposed method is useful in many cases for visualizing high-dimensional labeled data as well as for reducing the dimensionality for predictive model construction.
For visualization purposes, the proposed method appeared to perform better than PCA or SPCA, which is due to the algorithm's greedy nature that tries to maximally load all the predictive power on the first few features making it useful for exploratory analysis.
Regarding the predictive performance, although the method gave better results than the other methods on several problems, the experiments also demonstrated that it is not infallible and in some cases better results could be obtained by other means, such as the original SPCA.
Based on the results it seems that none of the considered dimension reduction techniques is optimal for every problem, but in almost all cases at least one of them gave very good results (clearly better than Lasso or ridge), confirming that the dimension reduction approach is very viable alternative for (Bayesian) supervised learning in these problems and encourages further research and methodological development in this area.
As it stands, our pragmatic advice would be to use cross-validation for assessing the fit of the models obtained after the different computational shortcuts (such as SPCA and ISPCA) and to use the validation results to guide the model selection. %
We emphasize that it is advisable to validate also the dimension reduction process (that is, the dimension reduction is computed separately for each fold) to avoid any potential bias induced by conditioning the inference twice on the observed data.

\newpage
{
\renewcommand{\section}[2]{}%
\subsubsection*{References}
\bibliographystyle{apalike}
\bibliography{references}
}

\clearpage
\section*{SUPPLEMENTARY MATERIAL}

\subsection*{Efficient Computation of the Projection Matrix}

Consider the computation of the ICPCA projection vectors $\vc w_k$, given by Equation~\eqref{eq:from_v_to_w}, which we repeat here for convenience
\begin{align}
\vc w_k = \left( \prod_{t=1}^{k-1} \vc A_t \right) \vc v_k %
		= \left( \prod_{t=1}^{k-1} (\vc I - \vc V_t \vc B_t) \right) \vc v_k.
\label{eq:from_v_to_w_repeated}
\end{align}
Recall that $\vc V_t$ has all columns equal to $\vc v_t$ (where $\vc v_t$ is computed at step 2 of iteration $t$) and $\vc B_t = \diag{\vc b_t}$ where $\vc b_t$ contains the coefficients $b_1,\dots,b_D$ from iteration $t$ (computed at step 3).

Consider now the first multiplication we need to compute in Equation~\eqref{eq:from_v_to_w_repeated}.
This can be rewritten as
\begin{align*}
\vc A_{k-1} \vc v_k
&= (\vc I - \vc V_{k-1} \vc B_{k-1}) \vc v_k \\
&= \vc v_k - \vc V_{k-1} \vc B_{k-1} \vc v_k  \\
&= \vc v_k - \vc v_{k-1} \vc 1^\tp \vc B_{k-1} \vc v_k \\ %
&= \vc v_k - \vc v_{k-1} \vc b_{k-1}^\tp \vc v_k \qquad | \quad c_{k-1} := \vc b_{k-1}^\tp \vc v_k \\
&= \vc v_k - c_{k-1} \vc v_{k-1}.
\end{align*}
Thus in order to compute the multiplication by matrix $\vc A_{k-1}$, all we need to do is to take an inner product between two vectors and then subtract two vectors which is very efficient.
To compute the full product \eqref{eq:from_v_to_w_repeated} we simply perform this operation in a loop, so that we first initialize $\vc v' = \vc v_k$ and repeat for $t=k-1, k-2,\dots, 1$ the operation $\vc v' = \vc A_t \vc v'$.
After the last multiplication we resulting vector will give us~$\vc w_k$.

\subsection*{Datasets}

The datasets we used for the comparisons are summarized in Table~\ref{tab:datasets}.
All of them are classification problems and most datasets are available at \url{http://featureselection.asu.edu/datasets.php}.
Although we are mostly interested in the ``small~$n$, large~$D$'' realm such as the microarray studies, we also wanted to consider how the different methods perform in other high-dimensional problems, such as in text classification where the features are typically word counts (Dexter, Basehock, PCMac).
Two of the datasets (Arcene, Dexter) are taken from the NIPS 2003 feature selection challenge (\url{http://clopinet.com/isabelle/Projects/NIPS2003/}).
These datasets are real problems but contain additional distractor features (probes) that have no predictive power.

\begin{table}[t]%
\centering
\abovetopsep=2pt
\caption{Summary of the real world classification datasets used for the experiments; dataset type, number of classes, dataset size $n$ and number of features $D$. Type `Gene' refers to gene expression data and `Text' to text classification (features are word counts). See supplementary material for more information.}
\label{tab:datasets}
\begin{tabular}{ lccccc }
\toprule
Dataset & Type & Classes & $n$ & $D$ \\ 
\midrule
Ovarian & Gene & 2 & 54 & 1536   \\
Colon & Gene & 2 & 62 & 2000   \\
Prostate & Gene & 2 & 102 & 5966   \\
Leukemia & Gene & 2 & 72 & 7129   \\
Glioma & Gene & 2 & 85 &  22283  \\
Glioma-4c & Gene & 4 & 50 & 4434  \\
Lung & Gene & 2 & 187 & 19993 \\
Lung-5c & Gene & 5 & 203 & 3312 \\
Arcene & Other & 2 & 200 & 10000  \\
Dexter & Text & 2 & 600 & 20000 \\
Basehock & Text & 2 & 1993 & 4862 \\
PCMac & Text & 2 & 1943 & 3289 \\
\bottomrule
\end{tabular}
\end{table}
\begin{figure*}[t]
	\centering
	\minput[pdf]{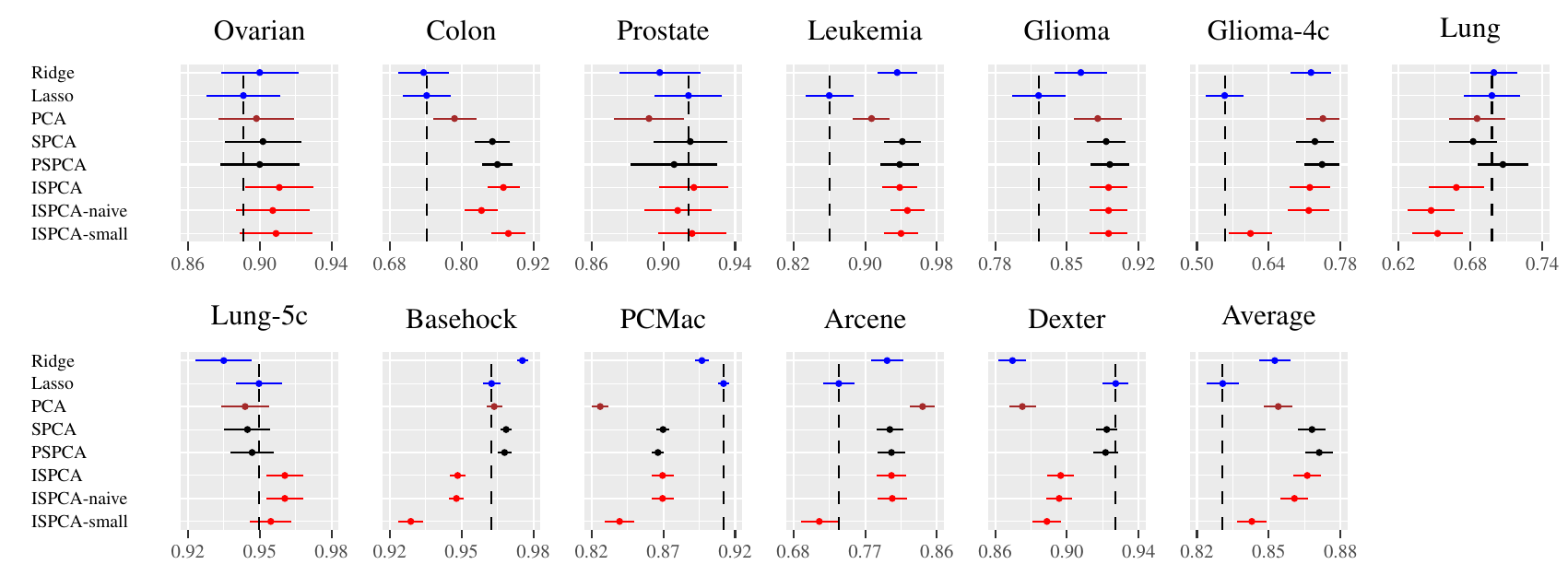}
	\captionspace
	\caption{Classification accuracies on test data for the different methods on different datasets (larger is better). Horizontal bars denote the 95\% intervals. The dashed vertical line denotes the performance estimate for the Lasso. The last plot denotes the average over all the datasets.}
	\label{fig:acc}
\end{figure*}
\begin{figure}[h!]
	\centering
	\minput[pdf]{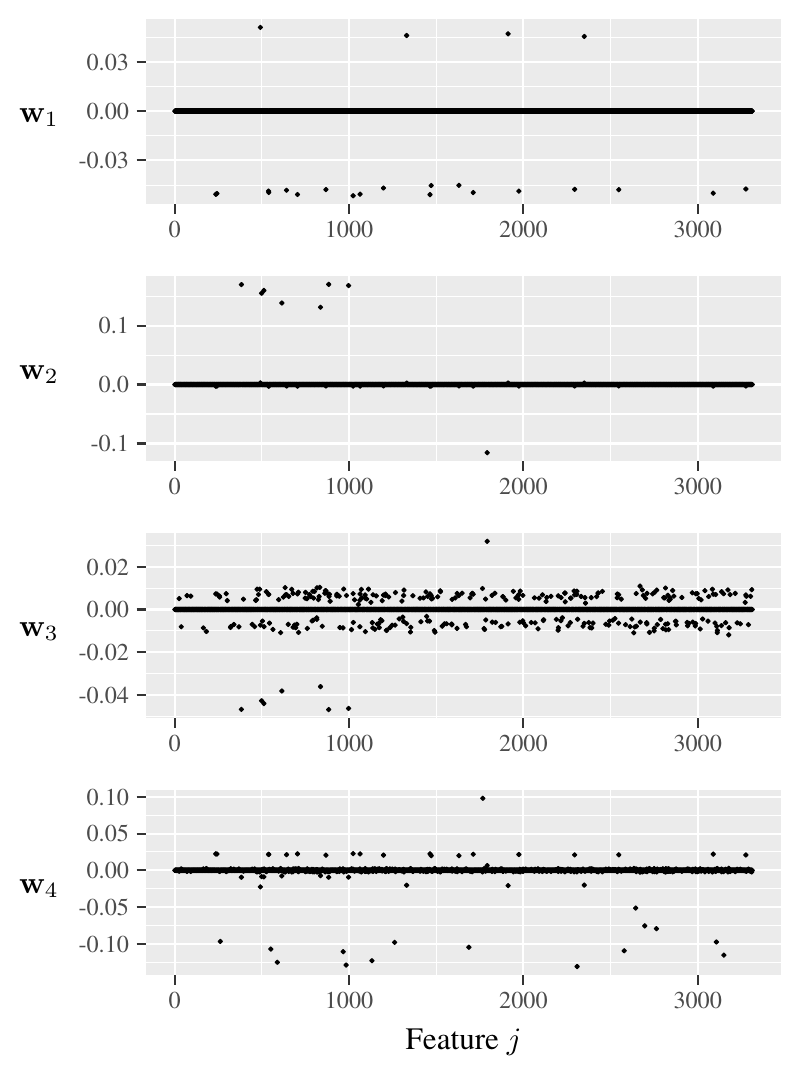}
	\captionspace
	\caption{The first four ISPCs (columns of the projection matrix $\vc W$) for the Lung-5c cancer data (${n=203, D=3312}$). The nonzero values indicate the genes that are characteristic for separating the corresponding class from the other classes (see Figure~\ref{fig:lung5c}).}
	\label{fig:lung5c_ws}
\end{figure}

\subsection*{Predictive Models and Priors}

In Section~\ref{sec:prediction}, for the binary classification problems we used standard logistic regression model
\begin{align*}
	p(y_i = 1 \given \vs \beta) = \frac{1}{1+\exp(-\vs \beta^\tp \vc x_i)},
\end{align*}
where $\vs \beta= (\beta_0,\beta_1,\dots,\beta_D)$ denotes the model parameters including the intercept $\beta_0$ (the notation assumes the first element of the predictor vector $\vc x$ is a constant $x_0=1$).
For the intercept we used a diffuse prior $\beta_0 \sim \Normal{0,10^2}$ and for the regression coefficients $j=1,\dots,D$ the regularized horseshoe \citep{piironen2017c} 
\begin{align*}
\begin{split}
	\beta_j \given \lambda_j,\tau,c &\sim \Normal{0, \tau^2 \ti \lambda_j^2},  \quad \ti \lambda_j^2 =  \frac{c^2 \lambda_j^2}{c^2 + \tau^2 \lambda_j^2}, \\
	\lambda_j &\sim \halfCauchy{0,1}, \\
	\tau &\sim \halfCauchy{0,\tau_0^2 }, \\
	c^2 &\sim \InvGamma{\nu/2, \nu s^2/2}.
\end{split}
\label{eq:rhs}
\end{align*}
This prior will shrink the coefficients of the irrelevant features heavily towards zero and softly regularize those that are far from zero.
Following the recommendations of the aforementioned paper, we chose $\tau_0=\frac{p_0}{D-p_0}\frac{2}{\sqrt{n}}$ with $p_0=1$ as our prior guess for the number of relevant features, and $\nu=4$ and $s=5$ as the parameters for the hyperprior on the regularizer $c^2$.

In the multiclass problems with $H$ classes we used the multinomial softmax regression
\begin{align*}
	p(y_i = \ell \given \vs \beta_1,\dots,\vs \beta_H ) = \frac{\exp(\vs \beta_\ell^\tp \,\vc x_i)} { \sum_{h=1}^H \exp(\vs \beta_h^\tp \,\vc x_i)}.
\end{align*}
We used the same prior as in the binary case, so that each of the $HD$ regression coefficients was given its own local scale parameter $\lambda_j$ with one global scale $\tau$.
This allows the regression coefficient for some feature to be far from zero for some class $h$ but be close to zero for the other classes, encoding the information that a feature can be relevant for separating one class from the others but irrelevant for separating the other classes from one another.

All the Bayesian models were fitted using Stan~\citep{stan_manual}, running 4 chains, 2000 samples each, first halves discarded as warm-up.
Ridge and Lasso solutions were computed with the default settings of the R-package {\tt glmnet} \citep{friedman2010}.

\subsection*{Extra Results}

Figure~\ref{fig:acc} shows the classification accuracies for the different models considered in Section~\ref{sec:prediction} and Table~\ref{tab:computation_times} typical computation times for some of the datasets.

Figure~\ref{fig:lung5c_ws} shows the first ISPCs for Lung-5c dataset considered for data visualization in Section~\ref{sec:visualization}.

\begin{table*}[t!]%
\centering
\abovetopsep=2pt
\caption{Average computation time (in seconds) over five repeated runs for a representative set of datasets. For PCA, SPCA and ISPCA, the time contains both the dimension reduction and model fitting (the number in the parenthesis indicating the relative amount of time spent in the dimension reduction), and for Lasso the cross-validation of the regularization parameter.}
\label{tab:computation_times}
\begin{tabular}{ lcccrrrr }
\toprule
Dataset & Classes &$n$ & $D$ & \multicolumn{4}{c}{Computation time} \\
\cmidrule(r){5-8}
& & & & PCA & SPCA & ISPCA & Lasso \\ 
\midrule
Leukemia & 2 & 72 & 7129  & 9.6 (2\%) & 8.3 (21\%) & 8.4 (24\%) & 1.0 \\
Glioma & 2 & 85 & 22283 & 14.6 (5\%) & 16.6 (33\%) & 14.5 (28\%) & 2.7 \\
Lung-5c & 5 & 203 & 3312 & 81.0 (1\%) & 82.2 (12\%) & 89.0 (19\%) & 5.2 \\
PCMac & 2 & 1943 & 3289 & 511.4 (2\%) & 303.3 (4\%) & 565 (22\%) & 18.9 \\
\bottomrule
\end{tabular}
\end{table*}
\FloatBarrier %
\phantom{fdsa}

\end{document}